\documentclass[10pt,conference]{IEEEtran}
\IEEEoverridecommandlockouts
\usepackage{cite}
\usepackage{amsmath,amssymb,amsfonts}
\usepackage{algorithm}
\usepackage{algpseudocode}
\usepackage{array}
\usepackage{subcaption}
\usepackage{multirow}
\usepackage{tabularx}
\usepackage{graphicx}
\usepackage{textcomp}
\usepackage{xcolor}
\usepackage{amsmath}
\usepackage{diagbox}
\newcommand{\CN}{\mathcal{CN}}
\def\BibTeX{{\rm B\kern-.05em{\sc i\kern-.025em b}\kern-.08em
    T\kern-.1667em\lower.7ex\hbox{E}\kern-.125emX}}

\def\Htran{\mbox{\tiny $\mathrm{H}$}}
\def\Ttran{\mbox{\tiny $\mathrm{T}$}}
\def\CN{\mathcal{N}_{\mathbb{C}}} 

 \newcommand{\vect}[1]{\mathbf{#1}}

\begin{document}

\title{Communicate or Sense? AP Mode Selection in mmWave Cell-Free Massive MIMO-ISAC 

\author{\IEEEauthorblockN{Weixian Yan\IEEEauthorrefmark{1}, Ozan Alp Topal\IEEEauthorrefmark{1}, Zinat Behdad\IEEEauthorrefmark{1}, \"Ozlem Tu\u{g}fe Demir\IEEEauthorrefmark{2}, and Cicek Cavdar\IEEEauthorrefmark{1}
\thanks{This work has been funded by Celtic-Next project RAI-6Green partly supported by Swedish funding agency Vinnova. The work by \"O. T. Demir was supported by 2232-B International Fellowship for Early Stage Researchers Programme funded by the Scientific and Technological Research Council of T\"urkiye.}
}  \IEEEauthorblockA{ \IEEEauthorrefmark{1} Department of Computer Science, KTH Royal Institute of Technology, Kista, Sweden}
\IEEEauthorblockA{ \IEEEauthorrefmark{2} Department of Electrical-Electronics Engineering, TOBB University of Economics and Technology, Ankara, Turkiye} 
\IEEEauthorblockA{
E-mail: (\{weixian, oatopal, zinatb, cavdar\}@kth.se, ozlemtugfedemir@etu.edu.tr)
		} 
}

\vspace{-4mm}
}
\maketitle

\begin{abstract}
Integrated sensing and communication (ISAC) is a promising technology for future mobile networks, enabling sensing applications to be performed by existing communication networks, consequently improving the system efficiency.  Millimeter wave (mmWave) signals provide high sensing resolution and high data rate but suffer from sensitivity to blockage. Cell-free massive multiple-input multiple-output (MIMO), with a large number of distributed access points (APs), can overcome this challenge by providing macro diversity against changing blockages and can save energy consumption by deactivating unfavorable APs. Thus, in this work, we propose a joint dynamic AP mode selection and power allocation scheme for mmWave cell-free massive MIMO-ISAC, where APs are assigned either as ISAC transmitters, sensing receivers, or shut down. Due to the large size of the original problem, we propose three different sub-optimal algorithms that minimize the number of active APs while guaranteeing the sensing and communication constraints. Numerical results demonstrate that assigning ISAC transmitters only satisfying communication constraints, followed up by sensing receiver assignment only for sensing constraint achieves the best performance-complexity balance.
\end{abstract}
\vspace{-2mm}
\section{Introduction}\label{section1}
\vspace{-2mm}

Integrated sensing and communication (ISAC) has emerged as a key technology for the sixth generation (6G) of wireless networks, offering sensing functionality seamlessly integrated into existing communication networks. This integration enables the network's sensing capabilities without additional hardware investment and enhances spectral efficiency via sensing-aided communications. Millimeter-wave (mmWave) bands ($30$-$100$\,GHz) can provide high sensing resolution and high data rate due to their wide bandwidth \cite{GhasrMohammadTayeb20133GLH}. To enable sensing, access points (APs) emit signals toward potential target locations and concurrently receive the reflected signals, namely monostatic sensing. Due to the sensitivity to blockage of the mmWave signals, the monostatic sensing can limit the performance of an AP if the location of the AP is unfavorable. 

One alternative solution is to utilize distributed APs that are capable of joint transmission/reception in the network, namely cell-free massive multiple-input multiple-output (MIMO), and select some APs as ISAC transmitters and the remaining as sensing receivers. However, allocating receive APs only for sensing may degrade communication performance. Furthermore, according to the target location and user equipments (UEs) in the system, the APs should dynamically adapt their modes to provide continuous high performance.  Thus, optimizing AP mode selection algorithms is essential to preserving the required system performance. 

In ISAC literature, the primary focus is on beamforming design and power allocation \cite{nguyen2023multiuser},
\cite{behdad2024multi}. Beyond the scope of ISAC, AP deployment problems are also addressed for indoor and outdoor communication scenarios in \cite{mmWave_1_TopalOzanAlp2023OJAP} and \cite{mmWave_2}, respectively. AP deactivation or shut-down is an approach providing significant reduction of the network power consumption considering the utilized processing and fronthauling resources \cite{behdad2024joint}.  In \cite{relatedwork-ming}, the base station (BS) mode selection in an ISAC cell-free massive MIMO network is studied, where a joint BS mode, beamforming design, and reception filtering method is proposed to maximize the sensing signal-to-interference-plus-noise ratio (SINR), while guaranteeing a certain communication rate to UEs. In the sensing part,  the authors consider multi-point target detection and, in the meantime assume perfect channel estimation in the communication part.  Besides BS mode selection,  a joint beamforming and mode selection framework is developed for the elements of reflector surfaces \cite{Zhang2024IntegratedSA}.

In this work,  we consider AP mode selection in mmWave cell-free massive MIMO-ISAC networks, aiming to minimize the total energy consumption of the network by reducing the number of active APs.  Different from \cite{relatedwork-ming}, we consider a passive target tracking in the sensing part, where the main performance indicator in our work for the sensing performance is Cram\'{e}r-Rao-lower-bound (CRLB). Furthermore, our methodology aims to minimize the number of active APs, while assigning some APs as ISAC transmitters and some APs as sensing receivers to guarantee communication and sensing constraints. Due to the large size of the considered problem, the global optimum is unattainable. We develop three different sub-optimal algorithms and compare their performance in terms of AP deactivation and the running time.

\vspace{-2mm}
\section{System Model}
\vspace{-2mm}

We consider a mmWave cell-free massive MIMO-ISAC system, consisting of $L$ APs, $K$ UEs, and one passive target with a predefined trajectory. Each AP has $N$ antenna elements, deployed in a horizontal uniform linear array (ULA). Communication and sensing signals are assumed to be transmitted in different frequency bands to eliminate any possible interference. After solving the proposed optimization algorithm, we decide the set of ISAC transmitters, $\mathcal{L}_{\mathrm{tx}}$, and set of sensing receivers, $\mathcal{L}_{\mathrm{rx}}$, where $|\mathcal{L}_{\mathrm{tx}}| = L_{t}$, and $|\mathcal{L}_{\mathrm{rx}}| = L_{r}$.

\vspace{-1mm}
\subsection{Communication Model}
\vspace{-1mm}

{\color{black}
We employ a distributed precoding approach, where APs independently decide their precoding based on the locally available channel state information, but they transmit the signal coherently.  We let $\varsigma_i\in \mathbb{C}$ denote the unit-power data signal of UE $i$,  $\mathbb{E} \{|\varsigma_i|^2 \} = 1$. We denote precoding vector and transmit power corresponding to UE $i$ and AP $l$ with $\vect{w}_{il}\in \mathbb{C}^N$ and $p_{il}\geq 0$, respectively. The transmit signal by AP $l$ is 
\vspace{-1mm}
\begin{equation} \vect{x}_l=\sum_{i=1}^K\sqrt{p_{il}}\,x_{i,l}\vect{w}_{il}\varsigma_i\in \mathbb{C}^{N}, \vspace{-1mm}
\end{equation}
where $x_{i,l}\in \{0,1\}$ equals to 1 if UE $i$ served by AP $l$ and 0 otherwise.
The received downlink signal at UE $k$ is
\begin{equation}  \label{eq:downlink-received-UEk}
y_k = \sum_{l\in \mathcal{L}_{\mathrm{tx}}}  {\vect{h}}_{kl}^{\Ttran} \vect{x}_l + n_k =\sum_{l\in \mathcal{L}_{\mathrm{tx}}} \sum_{i=1}^{K}\sqrt{p_{il}}  x_{i,l} {\vect{h}}_{kl}^{\Ttran} \vect{w}_{il} \varsigma_i + n_k 
\end{equation}
where $n_k \sim \CN(0,\sigma^2)$ is the receiver noise. $\vect{h}_{kl} \in \mathbb{C}^{N}$ is the channel vector obtained by ray tracing (RT) simulations as detailed in \cite{cfRT}. The downlink spectral efficiency (SE) of UE $k$ can be computed using \cite[Corr.~6.3 and Sec. 7.1.2]{cell-free-book} as 
	$\mathrm{SE}_{k} = \frac{\tau_d}{\tau_c} \log_2  \left( 1 + \mathrm{SINR}_{k}  \right)\,\textrm{bit/s/Hz}$
	with the effective SINR given by
\begin{equation}\label{eq:SINR-downlink-2}
\mathrm{SINR}_{k}=\frac{\left|\vect{d}_k^{\Ttran}{\boldsymbol{\rho}}_k\right|^2}{\sum\limits_{i=1}^K{\boldsymbol{\rho}}_i^{\Ttran}{\vect{C}}_{ki}{\boldsymbol{\rho}}_i+\sigma^2},
\vspace{-2mm}
\end{equation}
where $\tau_c$ and $\tau_d$ are the number of symbols and downlink data symbols in each coherence block, respectively and
\begin{align}
 &{\boldsymbol{\rho}}_k=\left [ \sqrt{p_{k1}}x_{k,1} \, \ldots \, \sqrt{p_{kL}} x_{k,L} \right ]^{\Ttran} \in \mathbb{R}_{\geq 0}^L,  \\
  &{\vect{d}}_k \in \mathbb{R}_{\geq 0}^L, \quad
    \left[ {\vect{d}}_{k} \right]_{l}= \mathbb{E} \left\{ \vect{h}_{kl}^{\Ttran}{\vect{w}}_{kl}\right\}, \quad {\vect{C}}_{ki}\in \mathbb{C}^{L \times L},  \\
  & \left[ {\vect{C}}_{ki} \right]_{lr}=\begin{cases}
  \mathbb{E} \left\{ \vect{h}_{kl}^{\Ttran}{\vect{w}}_{kl}{\vect{w}}_{kr}^{\Htran}\vect{h}_{kr}^*\right\}- \left[ {\vect{d}}_{k} \right]_{l} \left[ {\vect{d}}_{k} \right]_{r}^*, & i=k,  \\
  \mathbb{E} \left\{ \vect{h}_{kl}^{\Ttran}{\vect{w}}_{il}{\vect{w}}_{ir}^{\Htran}\vect{h}_{kr}^*\right\}, & i\neq k. \end{cases} 
\end{align}
Note that, during the downlink transmission phase, we assume the APs produce precoding vectors as they are serving all UEs. Later, if the optimization problem decides to turn off AP $l$, it will assign all transmit powers for AP $l$ to zero. We use a  modified version of the local partial minimum mean-square error (LP-MMSE) precoder, which can be found in \cite{ozlem_jsac}.

}
\vspace{-1.5mm}
\subsection{Sensing Model}
\vspace{-1.5mm}

We consider a multi-static sensing scenario, where some APs transmit the sensing signal, and some APs receive the reflected signal from the passive target. An extended target with the center of $(x,y)$ (in Cartesian coordinates) is being assumed. The system is tracking the target’s location and has available estimates for the target radar-cross section (RCS), from previous cycles. Transmit APs send orthogonal waveforms, $s_m(t)$, $\forall m$,  for the sensing \cite{CRB_power}. The signal transmitted from AP $m$ and  received by AP $n$ is given as
\begin{equation}
    \vect{y}_{m,n}(t) = \sqrt{\alpha_{m,n} \bar{p}_{m,s}}\beta_{m,n}\vect{H}^{\Ttran}_{m,n}\vect{w}_s s_m(t- \tau_{m,n}) + \zeta_{m,n}(t),
\end{equation}
where $\alpha_{m,n} \propto 1/(R^2_m R^2_n)$ is the channel gain, where $R_m = \|\mathcal{P}_T - \mathcal{P}_m\|^2$ is the range from transmitter AP (TX-AP) $m$ to the target, and $R_n = \|\mathcal{P}_T  - \mathcal{P}_n\|^2$ the range from the receiver AP (RX-AP) $n$ to the target. $\beta_{m,n}$ is the RCS coefficient.  $\bar{p}_{m,s}$ is the transmit power for sensing of AP $m$ and $\vect{w}_s$ is the sensing precoder. $\zeta_{m,n}(t)$ is circularly symmetric, zero-mean, complex Gaussian noise, spatially and temporally white with autocorrelation function $\sigma_{\zeta}^2 \delta(\tau)$. $\tau_{m,n}$ is the propagation delay, which can be calculated as 
\begin{equation}
    \tau_{m,n} = \frac{R_m +R_n}{c}.
\end{equation}
We assume an available line-of-sight (LOS) link between the APs and the target. The channel between the transmit AP $m$ and the receive AP $n$ can be given by
\begin{equation}
    \vect{H}_{m,n} \in \mathbb{C}^{N \times N} =  \vect{a}_m(\phi_m) \vect{a}^{\Ttran}_n(\phi_n),  
\end{equation}
where $\vect{a}_m(\phi_m)$ and $\vect{a}_n(\phi_n)$ are respectively the transmit and receive array vector responses with corresponding angle-of-departure and angle-of-arrival values. After receive combining, the equivalent received symbol is expressed as:
\begin{equation}
 \hat{y}_{m,n}\!(t)\! \!=\!  \!\sqrt{\!\alpha_{m,n} \bar{p}_{m,s}}\beta_{m,n}\!\vect{v}^{\Ttran}_n\vect{H}^{\Ttran}_{m,n}\!\vect{w}_s s_m(t\!-\! \tau_{m,n}) +\vect{v}^T_n \zeta_{m,n}(t),
\end{equation}
where $\vect{v}_n \in \mathbb{C}^N$ is the unit-norm receive combining vector.  We can abstract out the effect of beamforming by using an equivalent single-input single-output (SISO) channel as $\tilde{h}_{m,n} = \vect{v}^{\Ttran}_n\vect{H}^{\Ttran}_{m,n}\vect{w}_s$, where for the considered LOS case, $\tilde{h}_{m,n} = N$. 

We let $\vect{a} = [a_1, \ldots, a_L]^{\Ttran}$ denote the binary vector, where $a_l=1$ indicates that AP $l$ is chosen as ISAC transmitter. Similarly, we define $\vect{b} = [b_1, \ldots, b_L]^{\Ttran}$, where $b_l = 1$ indicates that AP $l$ is chosen as sensing  receiver. $a_l$ and $b_l$ cannot be simultaneously equal to $1$, as an AP cannot function as both TX-AP and RX-AP at the same time. We define  
\begin{subequations}
    \begin{align}
    [\vect{G}_{a}]_{l,l'}  = &\xi_l  \alpha_{l, l'}\left|\beta_{l, l'}\right|^2N^2\left(\frac{x_{l}-x}{R_{l}}+\frac{x_{l'}-x}{R_{l'}}\right)^2 \\  [\vect{G}_{b}]_{l,l'}  = &\xi_{l}  \alpha_{l, l'}\left|\beta_{l, l'}\right|^2N^2\left(\frac{y_{l}-y}{R_l}+\frac{y_{l'}-y}{R_{l'}}\right)^2 \\  
     [\vect{G}_{c}]_{l,l'} = & \xi_{l} \alpha_{l, l'}\left|\beta_{l, l'}\right|^2N^2\left(\frac{x_{l}-x}{R_{l}}+\frac{x_{l'}-x}{R_{l'}}\right)  \\ \nonumber
& \times\left(\frac{y_{l}-y}{R_{l}}+\frac{y_{l'}-y}{R_{l'}}\right), 
    \end{align}
\end{subequations}
where $\xi_l = \frac{8 \pi^2 B^2_s }{L^2 c^2 \sigma_{\zeta}^2}$, $B_s$ is the allocated bandwidth for sensing, and $c$ is the speed of light. 
The trace of CRLB can be interpreted as a lower bound of the sensing estimation error in cases when the sensing signal-to-noise ratio (SNR) is sufficiently high \cite{CRB_power}. Considering constant transmit power for sensing (i.e., $\bar{p}_{l,s} = P_s, \,\forall l$), we can write the trace of CRLB in terms of TX-RX mode vectors as 
\begin{equation}
    \operatorname{tr}(\bold{C}_t) = P^{-1}_s\frac{ \vect{a}^{\Ttran} (\vect{G}_a +\vect{G}_b)^{\Ttran}\vect{b}}{ \operatorname{tr}\left[ \left( \left(\vect{A}\vect{G}_b \right)^{\Ttran}\vect{G}_a - \left(\vect{A}\vect{G}_c \right)^{\Ttran}\vect{G}_c \right) \vect{B} \right]},
    \label{eq:CRLB}
\end{equation}
where $\vect{A} = \vect{a}\vect{a}^{\Ttran}$, and $\vect{B} = \vect{b}\vect{b}^{\Ttran}$.
\eqref{eq:CRLB} contains multiplicative terms of $a_l$ and $b_l$, which in the following section, we will reformulate to convexify the optimization problem.

\section{ISAC AP Mode Selection Problem}
The main aim of this work is to minimize the total energy consumption in the network, where the dominant energy consumption is resulted by the AP activation rather than the utilized transmit power \cite{behdad2024joint, mmWave_1_TopalOzanAlp2023OJAP}. In this case, the AP activation problem for a mmWave ISAC cell-free massive MIMO network is
\begin{subequations}
\begin{align}
  &\underset{p_{lk}, a_{l}, b_l}{\text{minimize}} \quad  \sum_{l=1}^L (a_l+b_l) \label{eq:optimization0:objective}\\
  & \text{subject to}  \nonumber \\
      & \frac{\left|\vect{d}_k^{\Ttran}{\boldsymbol{\rho}}_k\right|^2}{\sum\limits_{i=1}^K{\boldsymbol{\rho}}_i^{\Ttran}{\vect{C}}_{ki}{\boldsymbol{\rho}}_i+\sigma^2}    \geq  \gamma_{c}, \quad \forall k \label{eq:optimization0:SINR_constraint}  \\
      &  P^{-1}_s\frac{ \vect{a}^{\Ttran} (\vect{G}_a +\vect{G}_b)^{\Ttran}\vect{b}}{ \operatorname{tr}\left[ \left( \left(\vect{A}\vect{G}_b \right)^{\Ttran}\vect{G}_a - \left(\vect{A}\vect{G}_c \right)^{\Ttran}\vect{G}_c \right) \vect{B} \right]} \leq \eta \label{eq:optimization0:sensing_constraint}  \\
    & \sum_{k=1}^{K} p_{lk} \leq a_l P_{\mathrm{max}}, \quad \forall l  \label{eq:optimization0:total_transmit_power} \\
     & a_l + b_l \leq 1, \quad \forall l \label{eq:optimization0:AP_mode_selection} \\
    & a_l, b_l \in \{0,1\}\ \quad \forall l.\label{eq:optimization0:binary_constraint} 
\end{align}  
\label{eq:optimization0}
\end{subequations}
\eqref{eq:optimization0:SINR_constraint} is the SINR constraint of UEs, where $\gamma_c$ is the SINR threshold. 
\eqref{eq:optimization0:sensing_constraint} is the sensing constraint which is non-convex.  \eqref{eq:optimization0:total_transmit_power} limits the total transmit power ensuring that no power is allocated to an AP unless it is selected as an ISAC transmitter. \eqref{eq:optimization0:AP_mode_selection} ensures that an AP is either chosen as ISAC TX-AP, or sensing RX-AP, or shut down. \eqref{eq:optimization0:binary_constraint} makes the problem non-convex, and combinatorial due to the binary constraints. \eqref{eq:optimization0:SINR_constraint} and \eqref{eq:optimization0:sensing_constraint} are also non-convex in terms of the power coefficients and AP activation variables. We can reformulate \eqref{eq:optimization0:SINR_constraint}  into a second-order-cone (SOC) constraint as 
\begin{equation}
\left \| \begin{bmatrix} \sqrt{\gamma_c}\vect{C}_{k1}^{\frac{1}{2}}\boldsymbol{\rho}_1 & \ldots & \sqrt{\gamma_c}\vect{C}_{kK}^{\frac{1}{2}}\boldsymbol{\rho}_K &
\sqrt{\gamma_c}\sigma
\end{bmatrix} \right\|  \leq  \vect{d}_k^{\Ttran}{\boldsymbol{\rho}}_k,   \quad \forall k. \label{eq:Xconstraint1b}
\end{equation}
Although \eqref{eq:optimization0:sensing_constraint} can also be converted into a convex form by using binary varible transformations as in \cite{mmWave_1_TopalOzanAlp2023OJAP}, the problem size, in this case, requires $L^4$ number of binary variables, which is vastly high for a practical solution. Therefore, in the following, we will present three different solution methodologies to overcome this problem.

\subsection{Alternating Optimization}
One way to simplify \eqref{eq:optimization0:sensing_constraint} is to randomly choose a starting point for $\vect{a}$, and solve the problem to determine the sensing RX mode variables, $\vect{b}$. Then, the solution can be given as input to find the optimal ISAC TX-AP mode variables. To linearize \eqref{eq:optimization0:sensing_constraint},  $\vect{A} \in \mathbb{B}^{L\times L}$ can be defined as optimization variable, and $\vect{A} = \vect{a}\vect{a}^{\Ttran}$ can be guaranteed with the following constraints: 
\begin{subequations}
    \begin{align}
         &0 \leq	[\vect{A}]_{ij} \leq [\vect{A}]_{jj}, \forall i,j \\
 &0 \leq [\vect{A}]_{ii} - [\vect{A}]_{ij} \leq 1-[\vect{A}]_{jj}, \forall i,j  \\
     & [\vect{A}]_{ij} \in \{0,1\},  \quad \forall i,j 
    \end{align}
    \label{eq:binary_equvalentsA}
\end{subequations}

The optimization problem to find the ISAC transmitters for a given sensing receiver set is  
\vspace{-2mm}
\begin{subequations}
\begin{align}
  &\underset{p_{lk}, \vect{A}}{\text{minimize}} \quad  \operatorname{tr}(\vect{A}) \label{eq:optimizationalt:objective}\\
  & \text{subject to} \quad \eqref{eq:optimization0:total_transmit_power}, \eqref{eq:optimization0:AP_mode_selection} , \eqref{eq:Xconstraint1b}, \eqref{eq:binary_equvalentsA} \nonumber \\
      &  P^{-1}_s\frac{ \operatorname{diag}(\vect{A})^{\Ttran} (\vect{G}_a +\vect{G}_b)^{\Ttran}\vect{b}}{ \operatorname{tr}\left[ \left( \left(\vect{A}\vect{G}_b \right)^{\Ttran}\vect{G}_a - \left(\vect{A}\vect{G}_c \right)^{\Ttran}\vect{G}_c \right) \vect{B} \right]} \leq \eta.   
\end{align}  
\label{eq:optimizationalt}
\end{subequations}
This problem is in convex form except for the binary variables. Global optimum can be obtained by using a branch and bound algorithm. To linearize \eqref{eq:optimization0:sensing_constraint}, we can similarly introduce $\vect{B} \in \mathbb{B}^{L\times L}$ as an optimization variable, and $\vect{B} = \vect{b}\vect{b}^{\Ttran}$ can be guaranteed with 
\vspace{-2mm}
\begin{subequations}
    \begin{align}
         &0 \leq	[\vect{B}]_{ij} \leq [\vect{B}]_{jj}, \forall i,j \\
 &0 \leq [\vect{B}]_{ii} - [\vect{B}]_{ij} \leq 1-[\vect{B}]_{jj}, \forall i,j  \\
     & [\vect{B}]_{ij} \in \{0,1\},  \quad \forall i,j.
    \end{align}
    \label{eq:binary_equvalentsB}
\end{subequations}
After obtaining the solution, $\vect{A}$, we can solve the problem for only the sensing receivers. This problem can be given by 
\begin{subequations}
\begin{align}
  &\underset{\vect{B}}{\text{minimize}} \quad   \operatorname{tr}(\vect{B})\label{eq:optimizationaltb:objective}\\
  & \text{subject to} \quad  \eqref{eq:optimization0:AP_mode_selection} , \eqref{eq:binary_equvalentsB} \nonumber \\
      &  P^{-1}_s\frac{ \operatorname{diag}(\vect{A})^{\Ttran} (\vect{G}_a +\vect{G}_b)^{\Ttran}\operatorname{diag}(\vect{B})}{ \operatorname{tr}\left[ \left( \left(\vect{A}\vect{G}_b \right)^{\Ttran}\vect{G}_a - \left(\vect{A}\vect{G}_c \right)^{\Ttran}\vect{G}_c \right) \vect{B} \right]} \leq \eta.   
\end{align}  
\label{eq:optimizationaltb}
\end{subequations}
This problem is also non-convex and combinatorial due to the binary variables. However, the global optimum can be obtained by using a branch and bound algorithm since it is in a convex form except for the binary variables.  The algorithm for alternating optimization is given in Alg.~\ref{al:alternating}.

\begin{algorithm}[tb]
\caption{Alternation optimization algorithm}
\label{al:alternating}
\begin{algorithmic}[1] 
\State {\bf Input:} $\eta$: CRLB threshold, $\gamma_c$: SINR threshold, $\vect{G}_a$, $\vect{G}_b$, $\vect{G}_c$, $\vect{d}_k$, $\vect{C}_{ki}$.
\State {\bf Initializaton:} Initialize $\vect{b} \leftarrow \vect{b}^{(0)} $ with random binary vector. Set iteration counter to $i = 1$, convergence flag to $C = 0$.
\While{($C == 0$) \& ($i < I$)}
\State Check if  \eqref{eq:optimizationalt} is feasible.
    \If{\eqref{eq:optimizationalt} NOT feasible}
        \State Choose a new random $\vect{b} \leftarrow \vect{b}^{(0)} $.
    
    \Else
        \State Solve \eqref{eq:optimizationalt} with an input $\vect{b}^{(i-1)}$.
        \State Set  $\vect{a}^{(i)} \leftarrow \operatorname{diag} (\vect{A}^{(i)})$.
    \EndIf
    \State Solve \eqref{eq:optimizationaltb} with an input $\vect{a}^{(i)}$.
    \State Set  $\vect{b}^{(i)} \leftarrow \operatorname{diag} (\vect{B}^{(i)})$.
    \If{$(\vect{a}^{(i-1)}$ == $\vect{a}^{(i)})$ \& $(\vect{b}^{(i-1)}$ == $\vect{b}^{(i)})$}
        \State $C \leftarrow 1$.
        \State $i^{\star} = i$.
    \EndIf
    \State $i = i + 1$
\EndWhile
\State \textbf{Output:} $\vect{a}^{(i^{\star})}$, $\vect{b}^{(i^{\star})}$.
\end{algorithmic}
\end{algorithm}

Alg.~\ref{al:alternating} starts by randomly choosing sensing receivers. Then, it minimizes the number of ISAC TX-APs for this given random receiver set. After that, receiver activation vector is optimized for a given transmitter vector. This procedure iterates until the latest two TX and RX vectors are identical, demonstrating convergence to a solution. 
This algorithm is computationally inefficient for several reasons. First, it involves solving two mixed-integer optimization problems, both of which are NP-hard. Theoretically, this can result in very huge solution times for large-scale problems, unless early stopping is implemented, which comes at the cost of optimality gap. The second drawback is the sensitivity to the choice of a feasible starting point, as randomly guessing a feasible set often requires numerous attempts to solve the first problem. To address these challenges, we propose a simplification in the following section to reduce the computational complexity of this problem. 

\vspace{-1.5mm}

\subsection{Sequential Optimization}
\vspace{-1.5mm}

In this method, we first optimize the locations of ISAC TX-APs to only guarantee communication constraints, neglecting the sensing functionality. Then, we determine the sensing RX locations based on a given set of ISAC TX-AP locations. Since sensing is expected to be integrated into existing communication networks, this approach brings practical advantages to the existing network structure. The optimization algorithm is described and detailed in Alg. \ref{al:sequential}.
\begin{algorithm}[tb]
\caption{Sequential optimization algorithm}
\label{al:sequential}
\begin{algorithmic}[1]
    \State {\bf Input:} $\eta$: CRLB threshold, $\gamma_c$: SINR threshold, $\vect{G}_a$, $\vect{G}_b$, $\vect{G}_c$, $\vect{d}_k$, $\vect{C}_{ki}$.
    \State Solve the following problem with any mixed-integer solver: 
    \begin{subequations} 
    \label{eq:optimizationseq}
\begin{align}
  &\underset{p_{lk}, \vect{a}}{\text{minimize}} \quad  \sum_{l=1}^L(a_l) \label{eq:optimizationalt:objective}\\
  & \text{subject to} \quad \eqref{eq:optimization0:total_transmit_power},  \eqref{eq:Xconstraint1b} \nonumber \\
 & a_l \in \{0,1\}, \quad \forall l                  
\end{align}  
\end{subequations}
\State Set $\vect{a}$ to the solution of \eqref{eq:optimizationseq}.
    \State Solve \eqref{eq:optimizationaltb} with given  $\vect{a}$.
    \State \textbf{Output:} $\vect{a}$, $\vect{b}$
\end{algorithmic}
\end{algorithm}
Although Alg. \ref{al:sequential} is faster compared to Alg. \ref{al:alternating}, it still requires solving a mixed-integer problem. In the following, we propose a heuristic scheme, where we do not need mixed-integer optimization.

\subsection{A Heuristics Scheme}
The proposed heuristic scheme is given in Alg. \ref{alg:heuristic}. We first sort APs into two different lists. In the first list, we sort APs from highest to lowest channel gain. Here, the channel gain $g_{l}$ is the sum of the channel gains of the AP to all selected UEs, $g_{l}$ = $\sum_{k = 1}^K g_{lk}$. In the second list, we sort APs from closest to farthest from the target. We pick the first AP in the first list as an ISAC TX-AP. Next, we verify whether the SINR constraints of the UEs are met and continue adding APs until the constraints are satisfied. Then, similarly, we select the closest $R$ APs as sensing RX-AP. If the sensing threshold is not satisfied, then the highest channel gain AP is selected as ISAC TX-AP until the sensing threshold is satisfied. If the sensing threshold is satisfied, we gradually reduce the number of sensing RX-APs until the threshold is no longer met. At that point, we choose the last configuration, where the sensing threshold was still satisfied, as the final solution for the sensing RX-AP set.
\begin{algorithm}[tb]
\caption{Heuristics optimization algorithm}
\label{alg:heuristic}
\begin{algorithmic}[1]
\State {\bf Input:} $\eta$: CRLB threshold, $\gamma_c$: SINR threshold, Channel gain matrix, UE location, AP location.
\State {\bf Initialization:} Sort APs by $g_{l}$. Sort APs according to the distance between AP and target.
\State Add the AP which has the highest $g_{l}$ to $\vect{a}$, and closest $R$ APs to target to  $\vect{b}$.
\While{CRLB $> \eta$}
    \While{ \textbf{P1} NOT feasible}
        \State Select an available AP with the highest $g_{l}$ and add its index into $\vect{a}$.
        \State Solve $ \textbf{P1:} \, \underset{p_{lk}}{\text{min}} \,  \sum_{l=1}^L\sum_{k=1}^{K} p_{lk} \label{eq:check_tx_prob}
        \,\, \text{s.t.} \,\eqref{eq:optimization0:total_transmit_power},  \eqref{eq:Xconstraint1b} \nonumber $                  

        \State Check if \textbf{P1} is feasible
    \EndWhile
\EndWhile
\While{CRLB $< \eta$}
    \State Remove farthest RX-AP from $\vect{b}$.
\EndWhile
\State Add the last removed AP to $\vect{b}$.
\State \textbf{Output:} $\vect{a}$, $\vect{b}$.
\end{algorithmic}
\end{algorithm}

\section{Simulation Results}\label{results}
\begin{figure}
	\begin{minipage}[b]{0.48\columnwidth}
		\centering
		\subfloat[][]{\includegraphics[width=\linewidth]{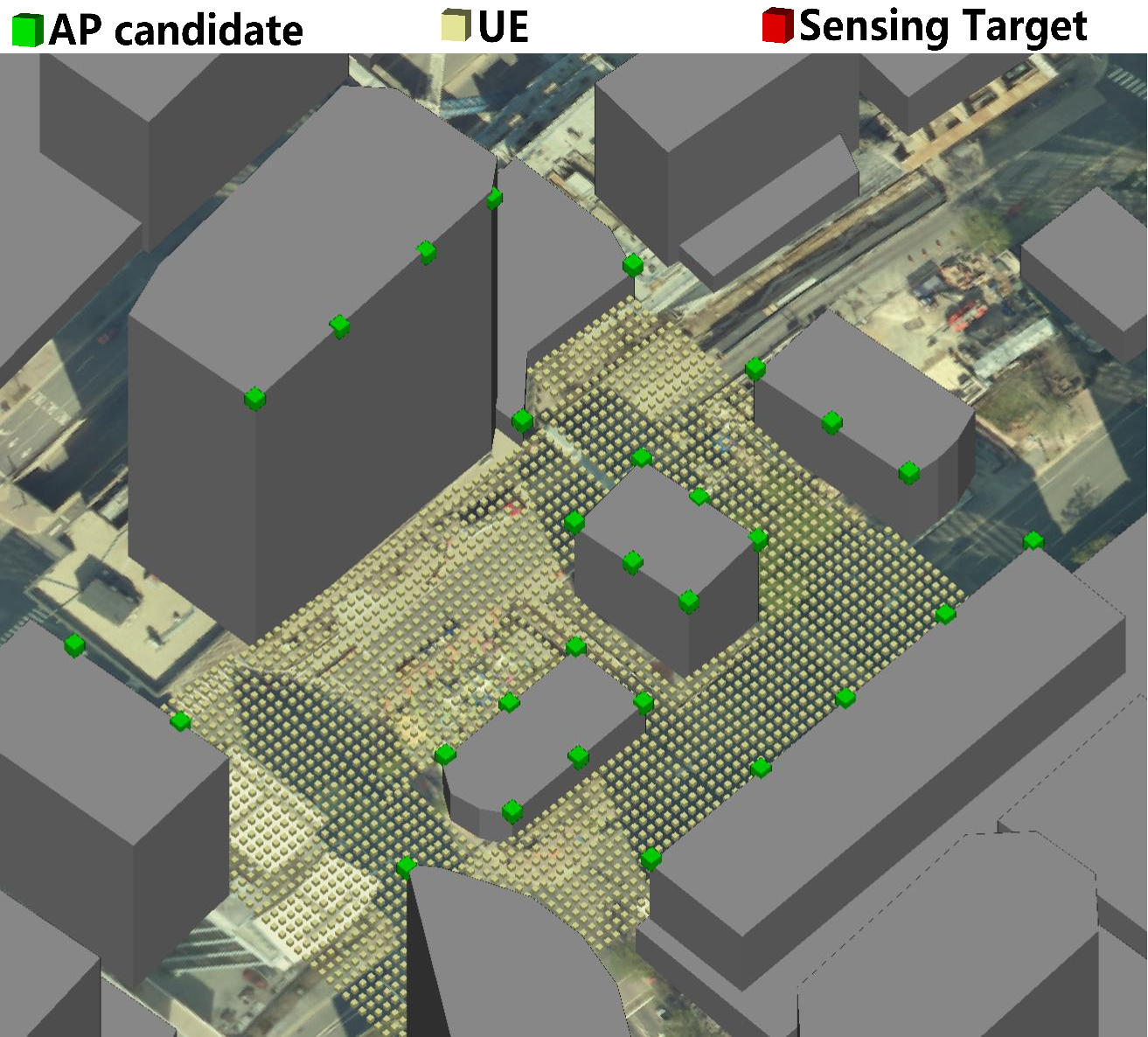}\label{fig:2.sub1}}
	\end{minipage}
	\begin{minipage}[b]{0.48\columnwidth}
		\centering
		\subfloat[][]{\includegraphics[width=\linewidth]{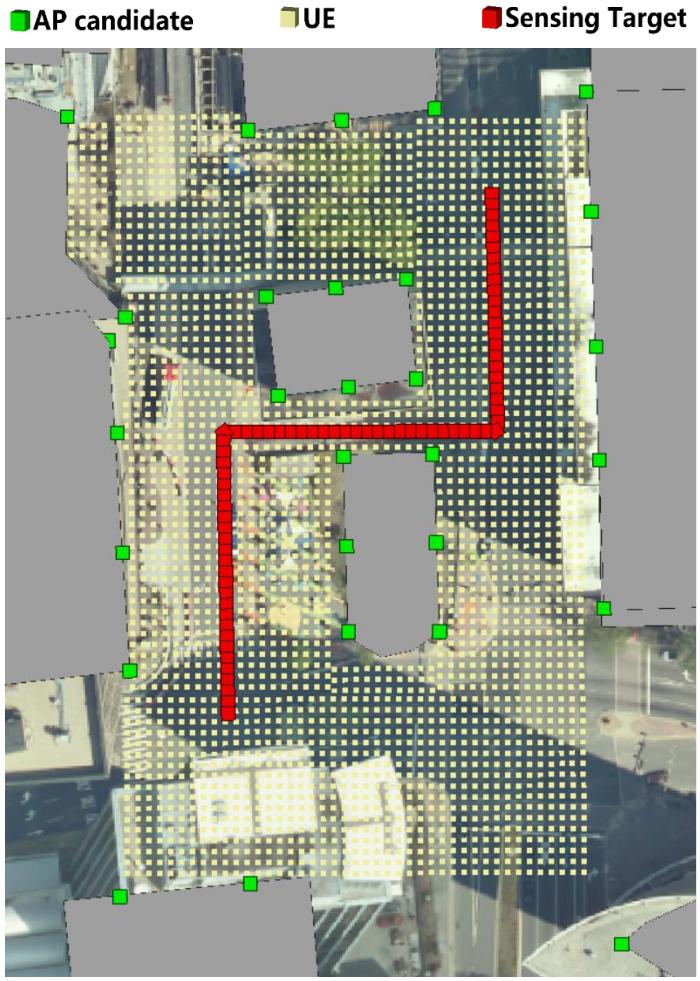}\label{fig:2.sub2}}
	\end{minipage}
	\caption{Candidate locations of APs and (a) UEs, (b) route of the sensing target.}
        \label{fig:Intro_3D_Total}
        \vspace{-4mm}
\end{figure}

We consider an outdoor urban area of Rosslyn, Virginia as shown in Fig.~\ref{fig:Intro_3D_Total}. A commercial RT tool, Wireless Insite, has been used for simulations. We set the number of APs to $L=12$. The highest channel gain-providing APs are selected from a set of $29$ candidate APs as shown in the green squares in Fig.~\ref{fig:2.sub1}. The carrier frequency is set to $28$ GHz, with all APs distributed along the top edges of buildings with the height of $10-100$\,m. We consider $2706$ candidate UE locations as in Fig.~\ref{fig:2.sub1}, and $207$ passive target locations, such as pedestrian or vehicle, with a predefined trajectory shown in Fig.~\ref{fig:2.sub2}. In the simulation experiments, 6 UE locations are randomly selected from all possible locations, while the target position is kept fixed. The TX-AP and RX-AP selection matrices are generated using the three optimization schemes for each setting. 300 Monte Carlo iterations are conducted for each setup, and the average numbers of TX-AP and RX-AP are calculated. 

\begin{figure}[t]
    \centering
    \includegraphics[width=0.8\linewidth]{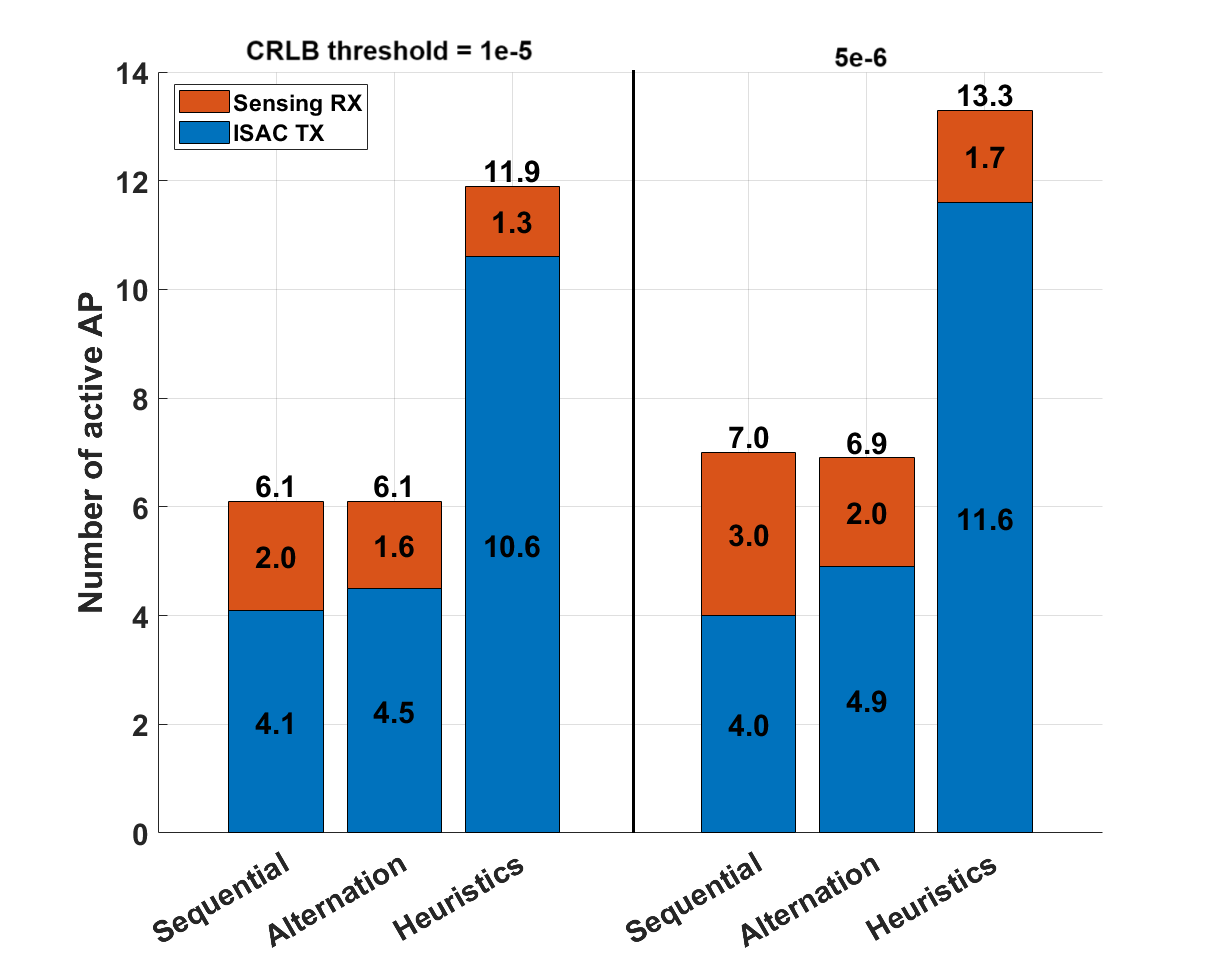}
    \caption{Performance of the proposed algorithms with communication SINR threshold of $20$\,dB and different CRLB thresholds.}
    \label{fig:same SINR diff CRB}\vspace{-6mm}
\end{figure}

Fig.~\ref{fig:same SINR diff CRB} shows the performance of the proposed algorithms under different sensing thresholds. The SINR threshold is fixed at 20 dB, while the CRLB threshold is set to $10^{-5}$ and $5 \times 10^{-6}$. It can be observed that optimization-based algorithms are more efficient in AP activation. Although both optimization-based schemes activate almost the same total number of APs, the sequential scheme can use fewer TX-APs, resulting in lower transmission power consumption.

\begin{table}\vspace{6mm}
    \centering
    \begin{tabular}{|c|c|c|c|}
         \hline
         $K$ & Sequential & Alternation
         & Heuristics\\
         \hline
         4 & 26 & 201.8 & 18.7\\
         \hline
         6 & 54.2 & 320.6 & 28.3\\
         \hline
         8 & 83 & 373.5 & 44.2\\
         \hline
    \end{tabular}
    \caption{Average runtime (in seconds) of three algorithms under different number of UEs.}
    \label{tab:run_time_table}
    \vspace{-4mm}
\end{table}

We compare the running time of the three optimization schemes as presented in Table \ref{tab:run_time_table}. The heuristics scheme has an absolute advantage in running speed because it does not use mixed integer optimization. Although the sequential and alternation schemes have similar optimization performance, the running time of alternation scheme is significantly longer— approximately eight times longer. This is because the alternation scheme undergoes multiple rounds of cyclic TX/RX optimization until convergence, substantially increasing computational complexity.

\vspace{-2mm}

\section{Conclusion} 

In this work, we propose an AP mode selection scheme for a mmWave cell-free massive MIMO-ISAC system. Deployed APs are dynamically assigned to one of three modes based on changing UE and target scenarios: (i) ISAC transmitter, responsible for transmitting joint sensing and communication signals, (ii) sensing receiver, or (iii) shut-down mode. The main goal is to minimize the number of active APs, thereby reducing the network energy consumption, while guaranteeing the communication and sensing constraints. Due to the non-convexity and huge size of the original problem, we propose three suboptimal algorithms, namely alternation optimization, sequential optimization, and heuristics. Sequential optimization achieves similar performance with alternation scheme, and both outperform the heuristics, by reducing the number of active APs by $50\%$. Additionally, the sequential algorithm offers approximately eight times shorter runtime compared to the alternation scheme, providing an effective balance between performance and computational complexity. 

\bibliographystyle{IEEEtran}
\bibliography{IEEEabrv,refs}
\end{document}